\documentclass[journal=ancac3,manuscript=article]{achemso}

\usepackage{chemformula} 
\usepackage[T1]{fontenc} 
 \usepackage{multirow}
 \usepackage{graphicx}
 \usepackage{amsmath}   
  \usepackage{booktabs}
\DeclareUnicodeCharacter{2212}{\textendash}
\usepackage{siunitx}

\author{Alexander Schröder}
\affiliation[University of Regensburg]
{Department of Physics, University of Regensburg, Regensburg, Germany}
\author{Andreas Wendeln}
\affiliation[University of Regensburg]
{Department of Physics, University of Regensburg, Regensburg, Germany}
\alsoaffiliation[Second University]
{Insitute of Physics, Carl-von-Ossietzky Universität Oldenburg, Oldenburg, Germany}
\author{Jonathan T. Weber}
\affiliation[University of Regensburg]
{Department of Physics, University of Regensburg, Regensburg, Germany}
\alsoaffiliation[Second University]
{Insitute of Physics, Carl-von-Ossietzky Universität Oldenburg, Oldenburg, Germany}
\author{Masaki Mukai}
\affiliation[JEOL]
{JEOL Ltd., Tokio, Japan}
\author{Yuji Kohno}
\affiliation[JEOL]
{JEOL Ltd., Tokio, Japan}
\author{Sascha Schäfer}
\affiliation[University of Regensburg]
{Department of Physics, University of Regensburg, Regensburg, Germany}
\alsoaffiliation[RUN]
{Regensburg Center for Ultrafast Nanoscopy (RUN), Regensburg, Germany} 
\email{sascha.schaefer@ur.de}

\title[An \textsf{achemso} demo]
{Laser-driven cold-field emission source for ultrafast transmission electron microscopy}
\abbreviations{IR,NMR,UV}

\begin{document}

\section{Abstract}
Ultrafast transmission electron microscopy (UTEM) has emerged as a versatile technique for the time-resolved imaging of nanoscale dynamics on timescales down to few-hundred attoseconds but the temporal and spatial resolutions are still limited by the coherence properties of pulsed electron sources. Here, we report the development of a novel laser-driven linear cold-field electron emitter integrated in a state-of-the-art UTEM system. Illuminating the sharp tungsten emitter tip with a UV light pulse generates ultrashort femtosecond electron pulses of 220 fs pulse duration, with energy widths as low as 360 meV. The photoelectron emitter demonstrates exceptional spatial coherence, achieving focal spot sizes down to 2 \AA ~and a peak normalized brightness exceeding 6.7~$\times 10^{13}$~A/m$^2$sr. With an order-of-magnitude improvement compared to previously employed laser-driven Schottky field emitters, the present development opens up the field of ultrafast atomic-scale electron probing.

\section{Introduction}
The rapid development of ultrafast transmission electron microscopy (UTEM) in recent years has widened the scope of this technique, now ranging from being a versatile tool for the time-resolved imaging of structural, electronic and spin degrees-of-freedom\cite{Barwick.2008,McKenna.2017,Kim.2020,Moller.2020,Cao.2021,Bucker.2016,Schliep.2017,Nabben.2023} as well as a platform for the detailed investigation of electron-light interactions on the nanoscale\cite{Vanacore.2018,Madan.2019,Henke.2021,Tsesses.2023}. In UTEM, the continuous electron beam of an electron microscope is replaced by femto- or picosecond electron pulses, which allows for the imaging of the instantaneous state of synchronized sample dynamics in a single-shot\cite{Lagrange.2008,Fu.2017,Picher.2018} or stroboscopic manner\cite{Flannigan.2012}.

The ultimately achievable performance in UTEM experiments in terms of spatial and temporal resolution is typically limited by the properties of the pulsed electron source. Important figures-of-merit include the number of electrons per pulse, the spectral bandwidth and -- crucial for microscopy applications -- the spatial coherence. In particular for high-resolution applications the electron beam brightness is a decisive quantity, connecting the emitted electron current with the emission area and the solid-angle into which electrons are emitted\cite{Reiser.2008,Rose.2012}.

Early designs for electron pulse generation relied on the photoemission from flat surfaces or truncated LaB$_6$ emitters triggered by femtosecond ultraviolet light pulses\cite{Barwick.2008}. With this approach, electron pulses with a large bunch charge can be obtained but only at a limited spatial coherence due to the large electron source area and significant Coulomb repulsion within the slowly accelerated electron pulse. 

Building on earlier work on multi-photon and strong-field photoemission from sharp metal tips\cite{Hommelhoff.2006, Ropers.2007}, more recent developments have introduced laser-driven Schottky field emitters in which photoemission is localized to the low-workfunction (100) front facet of a single-crystalline ZrO-covered tungsten tip\cite{Yang.2010,Feist.2017,Olshin.2020,Zhu.2020}. Utilizing this photoemitter approach, foci below 1~nm were demonstrated for continuous photoelectron beams, and also for femtosecond electron pulses, for which Coulomb correlations become important focal spot sizes in the range of a few nanometers were achieved,\cite{Bach.2019,Feist.2017,Olshin.2020} and applied in various ultrafast electron probing experiments \cite{Harvey.2020,Mattes.2024, Olshin.2021,Wang.2024}. Since the peak beam brightness of a laser-driven Schottky field emitter was found to be comparable\cite{Feist.2017} to the brightness of thermally emitted electron beams from Schottky sources\cite{Fransen.1999}, radio-frequency based beam choppers in combination with continuous Schottky sources could result in a sub-picosecond electron pulses with similar properties. Such RF chopper designs, based on RF cavities\cite{Verhoeven.2018} or traveling wave structures\cite{Fu.2020}, intrinsically operate in the GHz frequency range and, thus, these chopping approaches might be ideally suited for mapping sample dynamics with GHz excitation such as field-induced magnetic resonances.

Despite the substantial benefits of laser-driven Schottky field emitters over flat photocathodes in terms of beam brightness, Schottky field emitters also have some intrinsic disadvantages caused by the instability of the workfunction of the front facet due to a slow deterioration of the ZrO overlay. Depending on the vacuum quality around the emitter, replenishing the overlayer is necessary after some hours or days of operation and requires a prolonged heating to allow for fresh ZrO diffusing and segregating to the emitter surface.

Laser-driven cold-field emitters in ultrahigh vacuum (UHV) environments are a promising candidate as improved high-coherence pulsed electron sources. Operating such an emitter type in a linear photoemission regime\cite{Ehberger.2015,Haindl.2023} is highly desirable, to achieve sufficient photoemission current, to minimize the heat load on the nanoscale emitter apex and to allow for a broad tunability in electron pulse duration. However, up to now only a two-photon photoemission cold-field emitter source was implemented within a UTEM instrument\cite{Houdellier.2018}.

Here, we present the development of a linear, laser-driven cold field electron source integrated in an ultrafast transmission electron microscope. This approach yields femtosecond electron pulses with energy widths down to 360 meV and a peak normalized brightness exceeding 6.7 x 10$^{13}$ A/m$^2$sr. These findings underscore the potential of laser-driven cold field emitters to advance the UTEM technology by providing a new level of spatial and spectral precision in observing ultrafast nanoscale dynamics.
\begin{figure}[ht]
    \centering
    \includegraphics[width = 0.8\textwidth]{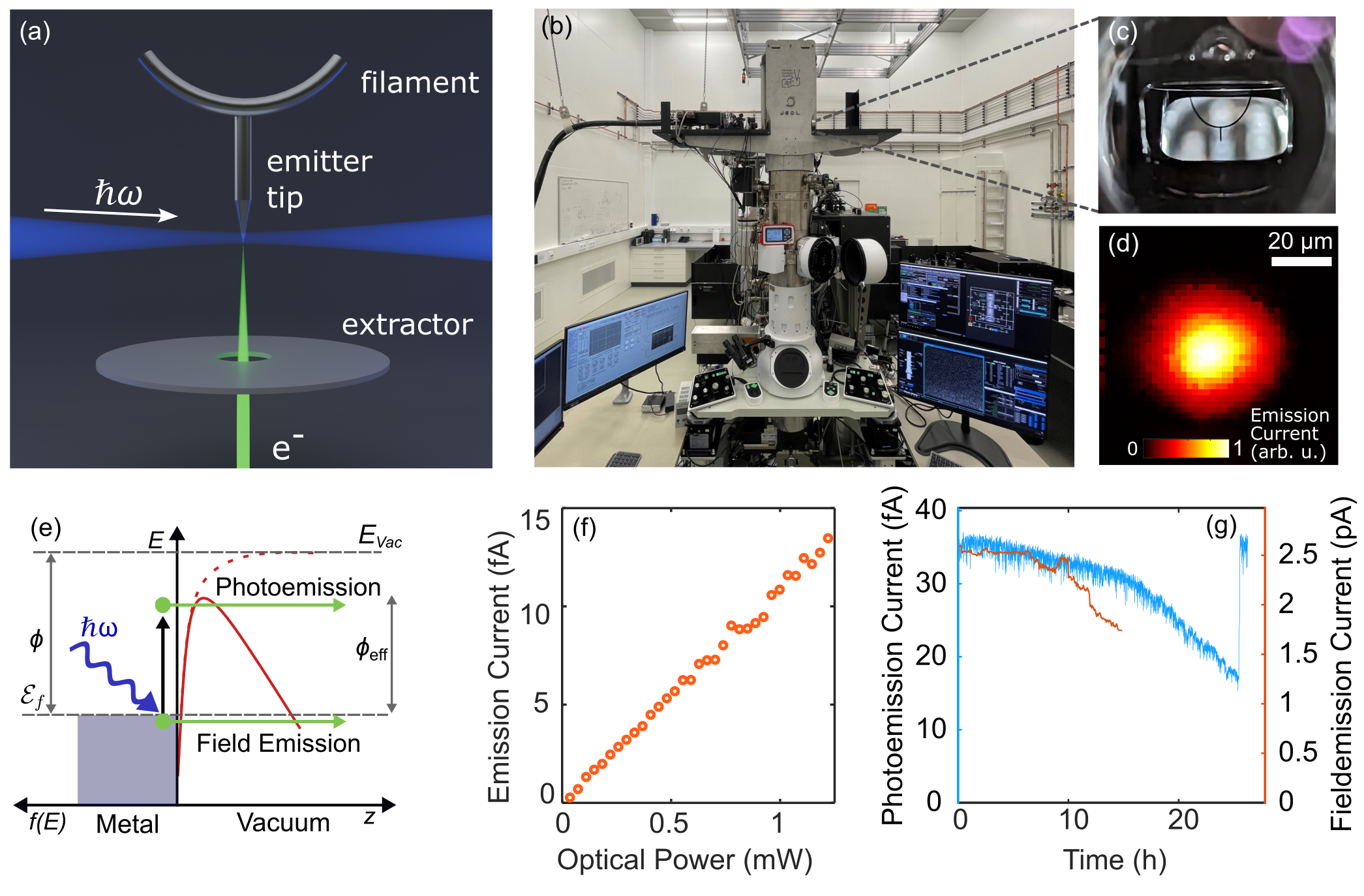}
    \caption{\textbf{Laser-driven linear cold field emission source}. (a) Schematics of the laser-driven electron emitter geometry. (b) Photograph of the modified cold-field emitter source mounted on the JEOL JEM F200 UTEM. (c) Photograph of the emitter tip seen through an optical window. (d) Electron emission map recorded by scanning the laser focus across the emitter tip and recording the current after the electron accelerator. (e) Schematic energy diagram for field- and photoemission processes. (f) Emission current as a function of optical power demonstrating a linear photoemission process. (g) Photoelectron emission stability (rolling averaged over 5 steps, blue curve) compared to the stability of the field-emission current (red curve). The photoemission current fully recovers after flash heating the emitter (here after 25 h).\label{fig:1}}
\end{figure}

\section{Results and Discussion}
The novel laser-driven cold field emitter is developed starting from the design of a commercial continuous cold-field emitter by JEOL Ltd.\cite{Sasaki.2010}. Briefly, a sharp single-crystalline (310)-oriented tungsten tip (radius of curvature about 70 nm) in an extreme high-vacuum (XHV, background pressure $<2 \cdot 10^{-11}$ mbar) is electrically biased by applying a potential difference between the tip and an opposing electrode, termed extractor (Fig. 1a). At high static extraction fields exceeding $F =4 \times 10^9$~V/nm, electrons tunnel through the workfunction barrier at the tip apex, resulting in a high brightness continuous electron beam which is further shaped by an additional electrostatic lens and accelerated to electron energies up to 200 keV. The modifications of the cold field-emitter electron source for a laser-driven operation mode include two opposing optical windows which have been added to the electron source allowing for a direct optical line-of-sight onto the tip (Fig. 1c). An optical assembly attached to the electron gun is used to focus optical pulses with varying photon energy $\hbar \omega$ on the emitter apex (200-mm focal length, focal spot size of about 20 µm, 0.12 maximum numerical aperture). Using the transmitted light at the exit window, the position of the laser focus on the emitter can be precisely monitored by the shape of the tip's optical shadow image. 

At the tip apex, the effective workfunction $\phi_{\mathrm{eff}}$ of the tip is locally reduced compared to the bulk value, $\phi$, due to the Schottky effect, allowing for a linear photoemission process if the photon energy of the illumination exceeds $\phi_{\mathrm{eff}}=\phi-\sqrt{\frac{e^3 F}{4\pi\varepsilon_0}}$.

At typical extraction fields applied for cold-field emission, the tunneling current (on the order of 10 µA) far exceeds achievable photoelectron currents. Lowering the applied extraction field to about 1.9 V/nm effectively closes the tunneling emission channel, while maintaining a reduced apex workfunction of about 2.6 eV. Thereby, illuminating the tip with a continuous wave (CW) laser with a wavelength of 355 nm (1-mW optical power, 3.49-eV photon energy, beam defocused on the emitter) indeed yields a photoemission current in the fA range as measured with a microchannel plate detector at the exit of the electron gun. An electron emission map is shown in Fig. 1(d), measured by moving the laser focus position across the emitter tip, and demonstrates that photoemission only occurs for direct illumination of the tip apex. The photocurrent scales linearly with the incident laser power (Fig. 1f), demonstrating a single-photon photoemission process. The photoemission yield is long-term stable, with no significant loss in photocurrent over an 8-hour period (Figure 1g, blue curve), comparable to or better than the stability of the field emitted electron current (Fig. 1g, red curve, measured at higher extraction fields).The photoemission yield drops to 50 \% on a 24-hour time scale and can be fully recovered by a short current flash (about 2-s duration) applied to a heating filament which holds the emitter tip. 

To further elucidate the dependence of field- and photoemitted currents on the applied extraction field, we systematically varied the extraction voltage and monitored the emitted total current at the exit of the electron source after the electron accelerator. Without tip illumination, the emitted current, $I$, closely follows a Fowler-Nordheim behavior, i.e. $I_{tunnel}/A=\frac{e^3F^2}{16\pi^2\hbar\phi}\exp\left(\frac{-4\sqrt{2 m_e}\phi^{3/2}}{3e\hbar F}\right)$ (Fig. 2a) with the electron mass $m_e$ and the emission area $A$ \cite{Fowler.1928}. Fitting the experimental data yields a $\beta_{\mathrm{tip}}$-factor of 1.8 x $10^6$ m$^{-1}$, which determines the relation between the applied extraction voltage, $U_{\mathrm{extr}}$,  and the electric field at the tip apex $F=\beta_{\mathrm{tip}}U_{\mathrm{extr}}$. Specifically, at an applied relative extractor potential of about 1.3 kV, we obtain an extraction field of 2.34 x $10^9 \mathrm{V}/\mathrm{m}$. We note that over the course of months of operation, applying daily current flashes to the tip results in a gradual change of the tip's radius of curvature and thus a slow drift of $\beta_{\mathrm{tip}}$.

\begin{figure}[t]
    \centering
    \includegraphics[width = 1\textwidth]{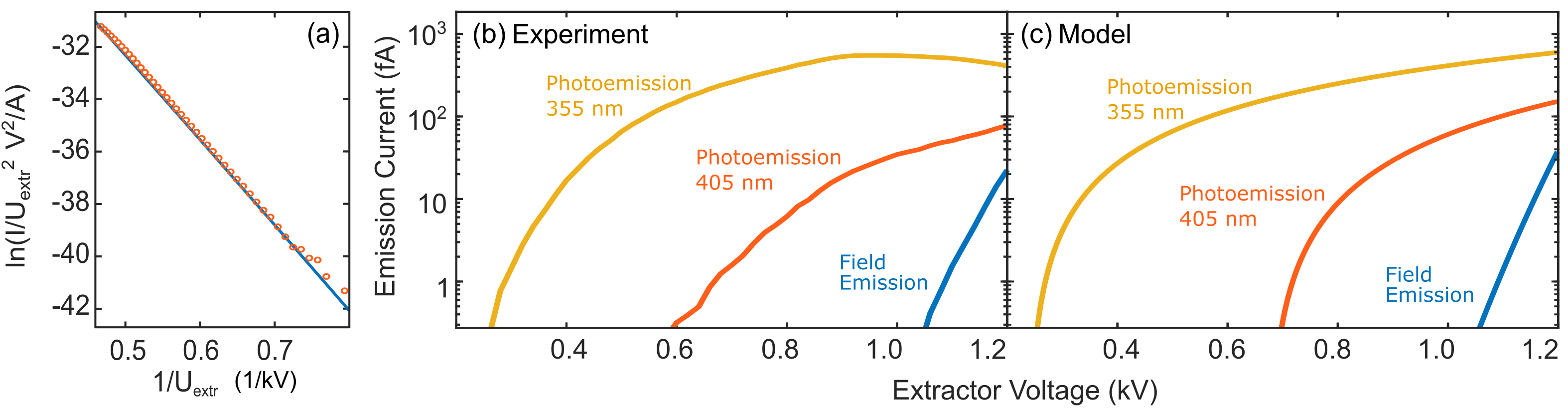}
    \caption{ \textbf{Current characteristics of the field-localized photoelectron source}. (a) Fowler-Nordheim plot of cold field emission (orange circles) and theoretical emission model at a $\beta_{\mathrm{tip}}$ factor of 1.8 x $10^6$ m$^{-1}$ (blue line). (b,c) Total electron CW-photoemission current at 1-mW optical power as a function of extraction voltage and comparison with the emission model.\label{fig:2}}
\end{figure}

Illuminating the tip with a CW optical beam at 355- and 405-nm wavelength, respectively, and 1-mW optical power yields a photoelectron current already at lower extraction field values, as shown in Fig. 2b. Whereas below an extraction voltage of about 1.1 kV, the tunnel emission current is below the noise floor of the detection, the photocurrents for both employed wavelengths show lower voltage thresholds at 0.3 kV (for 355 nm) and 0.6 kV (405 nm), respectively. Remarkably for the 355-nm case and at an optimized extraction voltage of about 1 kV, the photoemission current exceeds 400~fA at the exit of the source. Due to beam-defining apertures in the extractor electrode and at the source exit, a substantial portion of the emitted electrons at the emitter is filtered out. By comparing the emission current with the detected electrons at the exit, we estimate a maximum transmission of about 3\% in an electrostatic lens setting optimized for transmission. Therefore, the maximum observed photocurrent at the tip is about 13 pA (at 1-mW optical excitation.

The field-dependence of the photocurrent can be quantitatively reproduced by a photoemission model originally developed for flat photocathodes\cite{Dowell.2009} and employing a field-dependent effective workfunction (Fig. 2c, see supporting information for details). Utilizing an estimated emission area of 150 nm$^2$, a parabolic band model for tungsten\cite{Mattheiss.1965}, and an electron-optical transmission of about 3\%, results in a remarkable semi-quantitative agreement with the experimental data for the three data sets and without employing further adjustable parameters.

In our approach, the onset of photoemission occurs when the photon energy equals the effective work function, i.e. $U_{\mathrm{thres}}=4\pi\epsilon_0/(\beta_{\mathrm{tip}} e^3)(\phi-\hbar\omega)^2$. At higher extraction fields, the number of electronic states from which photoemission can occur increases, resulting in the experimentally and theoretically observed photocurrent increase. Due to the employed parabolic band model, the predicted above-threshold photocurrent scales with $I_p\propto\sqrt{U_{\mathrm{ext}}-U_{\mathrm{thres}}}$. Some deviations between experimental data and theoretical predictions are visible in the high-current regime, in particular for illumination with a 355-nm wavelength for which a local maximum in the current is observed at 0.95 kV (Fig. 2b, yellow curve). Such behavior might be attributed to a more complicated structure in the density-of-states of tungsten as well as to potential high-energy electronic resonances close-to or above the workfunction. We note that although an operation in this high-current regime might be useful in some cases, the expected higher excess electron energy after emission will likely increase the effective emitter source size and the obtained beam brightness.
\begin{figure}[t]
    \centering
    \includegraphics[width = 1\textwidth]{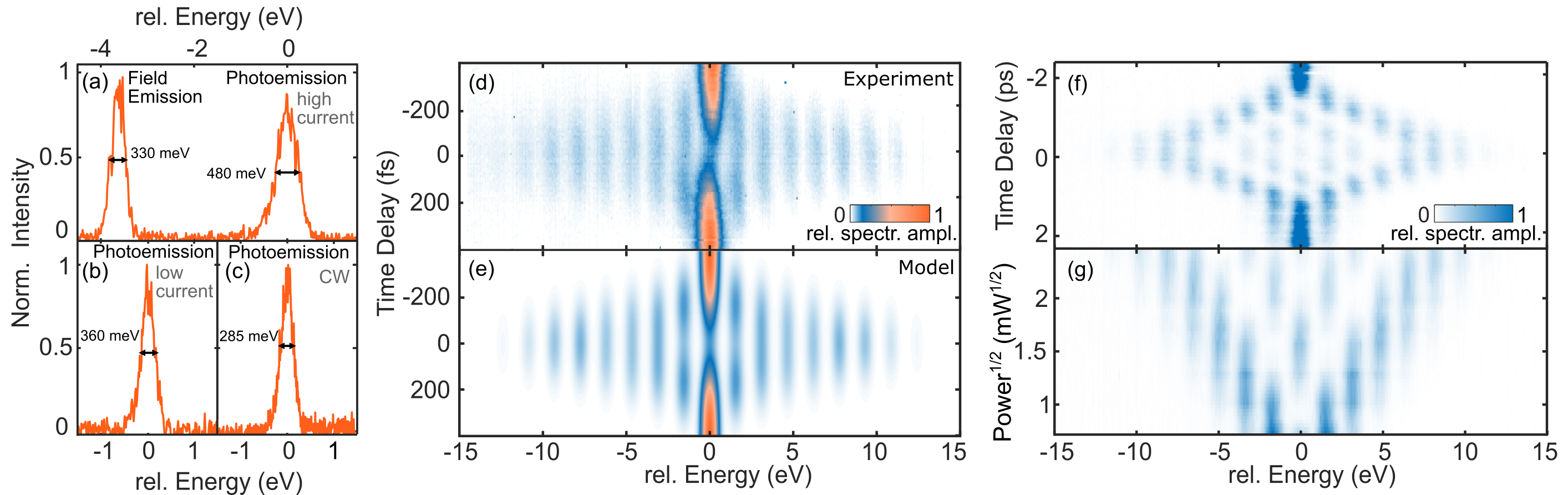}
    \caption{\textbf{Longitudinal femtosecond electron pulse properties.} (a-c) Electron energy width for different electron beam settings. (a) The extraction field is adjusted to equally enable continuous cold field emission and fs-photoemssion. At this setting the photoemission energy width is due to space charge effects. Lower extraction fields in (b) result in a narrower energy distribution of 360 meV. With CW-photoemission (c) the distribution narrows to 285 meV. (d) Electron energy loss spectra as function of delay in the presence of a 169 fs laser pulse. Reproducing the electron-photon cross-correlation with a numerical simulation yield a pulse duration of 220 fs. (f) Electron energy loss spectrum for a longer laser pulse duration of 1.6 ps. (g) Electron energy spectrogram at a zero delay as a function of the square root of the incident optical power (linearly interpolated in y-direction). \label{fig:3}}
\end{figure}

For further characterization of the transverse and longitudinal photoelectron pulse properties, we mount the photoelectron source on the Regensburg ultrafast TEM which is based on a modified JEOL JEM F200 microscope with a post-column CEOS imaging energy filter (Fig. 1(b) ). Choosing a femtosecond optical illumination of the emitter (343-nm wavelength, 0.25-mW optical power, 400-kHz repetition rate) and an extraction field for which photo- and tunnel-emitted currents are roughly equal, the recorded electron energy spectrum (Fig. 3a) consists of two peaks separated by the employed photon energy of 3.6~eV. Considering the energy diagram in Fig. 1c, the energy difference between both emission channels is expected since the field-emitted electrons from close to the tungsten's Fermi energy $\mathcal{E}_f$ tunnel through the barrier, whereas photoemitted electrons are born at an energy of $\mathcal{E}_f+\hbar\omega$. The minimum spectral width of the field-emitted electrons is related to the energy-dependent tunnel-efficiency through the barrier\cite{Zuo.2017, Swanson.2009} and further experimental fluctuations, and reaches in our case about 320~meV. The spectral width of the femtosecond photoelectron pulse is slightly enlarged to about 480~meV due to space charge repulsion within the confined electron bunch\cite{Bach.2019}. Reducing the optical power on the tip and the applied extraction field, the spectral width of the femtosecond photoelectron pulses can be further narrowed down to 360~meV (Fig. 3b). By using CW illumination, electron-electron Coulomb repulsion can be minimized, which results in a photoelectron energy width of 285~meV (Fig. 3c). Notably, CW field- and pulsed photoelectron emission are only occurring simultaneously for extractor voltages exceeding the tunnel-emission onset. For most ultrafast TEM experiments, a lower extraction field would be chosen (as employed for Fig. 3(b,c)) so that no field-emitted electron current contributes to the detected signal.

We determine the electron pulse duration by electron-photon cross-correlation in which electrons are passing the optical near-field of an illuminated aluminum thin film and experience inelastic electron-light scattering, also known as photon-induced near-field electron microscopy (PINEM)\cite{Barwick.2009,Park.2010,GarciadeAbajo.2010,Vanacore.2019b,Piazza.2015,Wang.2020,Gaida.2024,Muller.2024}. The recorded electron energy spectra as a function of electron-light delay are plotted in Fig. 3(d) and exhibit several photon-sidebands on the energy-gain and loss side of the spectrum. Taking into account the optical pump pulse parameters (see supporting information for details), the electron-light spectrogram can be well reproduced by considering an electron pulse duration of 220~fs (Fig. 3(e)). With the optical pulses stretched to about 1.6~ps, all electrons within a temporal pulse distribution experience an approximately constant light field amplitude when interacting with the light field. The corresponding spectrogram shows a distinct modulation of the individual photon sideband amplitudes (Fig 3f), attributed to the inherent coherence of the scattering process. For increasing light fields, coherent Rabi oscillations with a near-total depopulation of the initial energy state are observed (Fig. 3(g))\cite{Park.2010,Feist.2015}, highlighting the achievable spatial and temporal coherence of the electron probing.

The transverse beam properties of electron pulses are crucial for ultrafast imaging, local probing, and diffraction applications. A key metric for the beam quality is the normalized beam emittance  $\varepsilon_{n,x/y}$ in the transverse x- and y-directions, for which the product $\varepsilon_{n,x}\cdot\varepsilon_{n,y}$ describes the conserved occupied phase space volume of the electron beam \cite{Reiser.2008}, and is related to the relative transverse degree-of-coherence $K=\xi_c/\sigma_x=\hbar/m_{ec}\cdot1/\varepsilon_{rms_x}$ as well as the beam quality factor $M^2=1/K$. In a non-aberrated focal spot, the beam emittance is given by $\varepsilon_n,x=\beta\gamma\sigma_x\sigma_{\alpha x}$ in which $\sigma_x$ is the root-mean-square value of the spatial electron density distribution along the x-direction and $\sigma_{\alpha x}$ the corresponding value for the angular distribution. The constants $\beta$ and $\gamma$ are introduced to normalize for the improved coherence at higher electron energies and are given by $\beta=v_e/c$ with the electron velocity $v_e$ and the speed of light $c$ and the Lorentz factor $\gamma=1/\sqrt{1-\beta^2}$. Since the emittance can be improved, at the expense of electron current by aperturing the beam, the normalized beam brightness $B_n=I/(4 \pi^2\varepsilon_{n,rms}^2)$ is considered a prime figure-of-merit for electron sources. We note that in the electron microscopy community often a related quantity, the reduced brightness $B_r\propto B_n\beta^2\gamma^2/U_\textrm{acc}$ is used, in which $U_\textrm{acc}$ is the acceleration voltage. The proportionality constant depends on the adopted definitions of the angular and spatial spread\cite{Kruit.2016}.

In order to determine the brightness of the laser-driven cold-field emitter, we measured for an electron energy of 200 keV the spatial and angular distributions of focused photoelectron beams for different condenser parameters, including condenser aperture (CLA) diameter, spot-size- and angular-settings and different condenser modes ("TEM"/"Probe"). As fs-electron pulses are susceptible to mean-field and stochastic Coulomb effects leading to an increased beam emittance\cite{Bach.2019}, we first characterize the transverse beam properties for continuous wave (CW) photoemission (illumination with 405-nm wavelength, 25-mW optical power). An example for the observed beam profiles in the sample plane for a 40-µm CLA, spot size 5, angle 5 in probe mode (abbreviated in the following as Probe/5/5/40) is shown in Fig. 4(a), achieving a sub-nanometer photoelectron probe with a spatial width of only 2.1 \AA~(full-width-at-half-maximum, FWHM), a semi-convergence angle of 4.11 mrad and a current of 630 aA. From these measurements, we obtain a normalized emittance value of $\varepsilon_n=0.36$ nm$\cdot$mrad and an average normalized beam brightness of $B_n=1.25\times10^8$ A/m$^2$sr. Remarkably, the degree of coherence $K=\xi_c/\sigma_x$, with the transverse coherence length $\xi_c=\hbar/m_{ec}\cdot \sigma_x/\varepsilon_{rms_x}$, reaches up to 54\% (further beam characterizations are available in the supporting information).

\begin{figure}[]
    \centering
    \includegraphics[width = 0.8\textwidth]{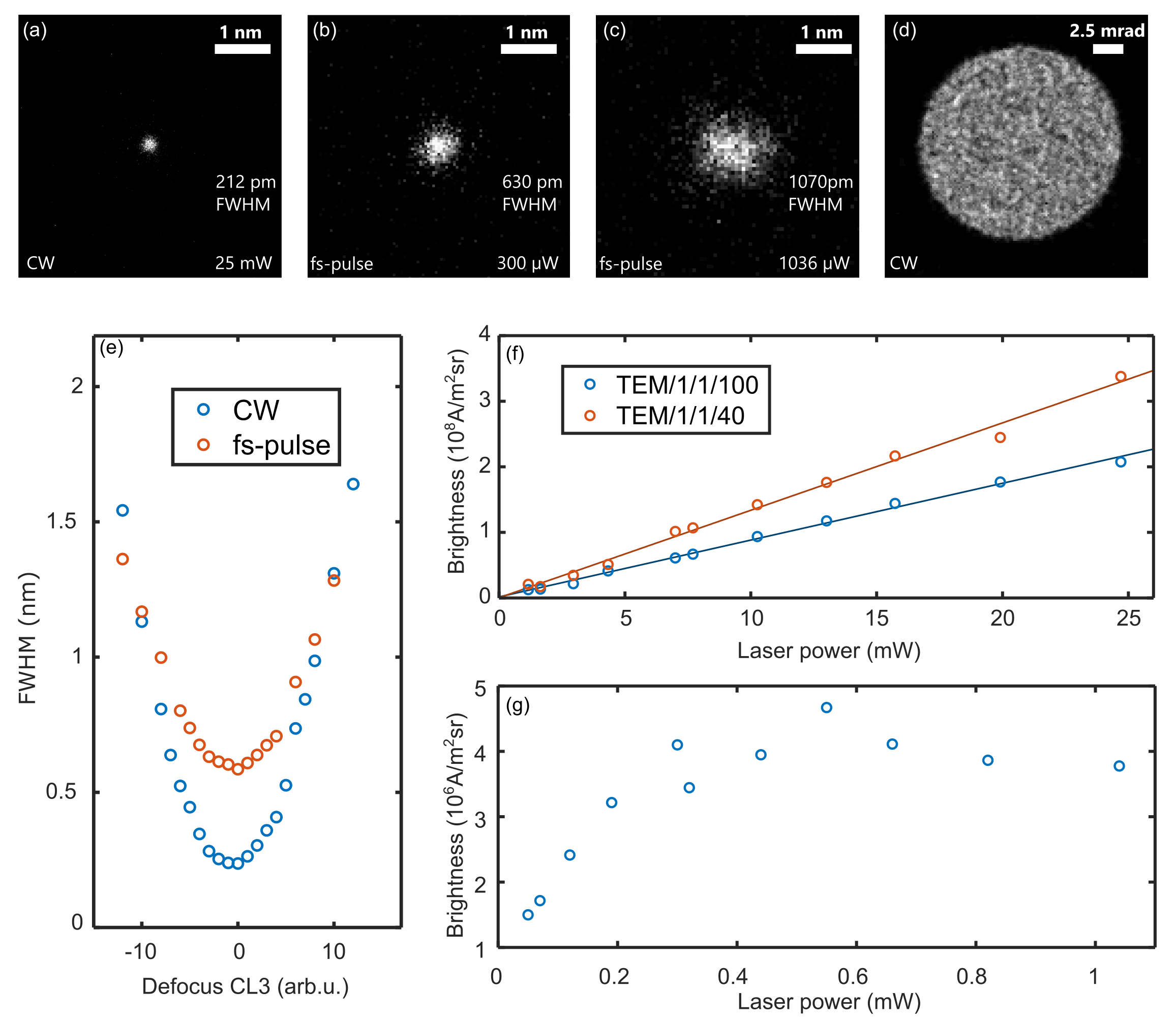}
    \caption{\textbf{Transverse femtosecond electron pulse properties.} (a-c) Electron micrographs of the electron focal spots for a Probe/5/5/40 condenser setting and CW optical illumination at 25 mW incident power (a) and femtosecond illumination at 300 µW (b) and 1036 µW (c) power, respectively. (d) Angular distribution of photoelectron beams shown in (a-c). (e) Electron beam caustics for CW-photoelectron beam and fs-electron pulses shown in (a,b). (f-g) Electron beam brightness as a function of optical power for both CW-photoemission (f) and fs-electron pulses (g).\label{fig:2}}
\end{figure}

Considering the case of femtosecond electron pulses, we estimate an emitted peak current of up to 110 µA, considering the 3\%-transmission of the gun, a 400-fA photocurrent maximum observed after the gun at 1-mW optical power and the optical pulse length of 300 fs. At such high instantaneous currents, Coulomb interactions within the pulse are expected. At a reduced optical power of 300 µW (0.3 nJ pulse energy, 343 nm central wavelength), we find electron focal spot sizes considerably increased compared to the CW case, as visible in the comparison between Fig. 4 (a) and (c), indicating stochastic Coulomb interactions within the electron pulse for larger optical powers \cite{Bach.2019}. The experimental power-dependence of the normalized beam brightness is shown in Fig. 4(g), further highlighting the initial linear increase at small currents and the onset of Coulomb interactions and the saturation of beam brightness at about 0.3 mW. Notably, for the current experiments, the low-power limit of the brightness for femtosecond optical illumination does not need to coincide with the CW case due to the different photoemission wavelengths. 

We note that the average current, and thus the average brightness, for a given UV optical pulse energy on the emitter depends on the repetition rate of the experiment. The 400-kHz repetition rate adopted here is well suited for investigating optically-driven lattice and spin dynamics in many sample systems. Investigations of electron-near-field scattering and radio-frequency-driven dynamics can be conducted at higher repetition rates such as 80 MHz, allowing for average currents in sub-nanometer femtosecond pulses exceeding 6.25 fA for an average incident optical power of 25 mW (as used in the CW case). 

To account for the effective duty cycle $D_{eff}=\sqrt{2}f_{rep}t/\sqrt{8 ln(2)}$ of pulsed emitters, the normalized peak brightness  B$_{np}=B_n/D_{eff}$ allows for comparison with continuous electron sources. With a slightly space charge broadened electron pulse duration of $\Delta t=300$ fs and a repetition rate $f_{rep}$ = 400 kHz, this corresponds to a normalized peak brightness B$_{np}$ of about $6.7\times 10^{13}$ A/m$^2$sr, similar to continuous cold field emission sources.

Whereas the performance of a pulsed photoemitter depends on many instrumental parameters, such as optical illumination conditions, optimization of the coupling between emitter and electron optics and details of the emitter geometry, it is instructive to compare published brightness measurements of laser-driven Schottky emitters\cite{Feist.2017,Olshin.2020}, indicating that the smaller emission area and larger electric extraction field of a laser-driven cold-driven enables an improvement in the average beam brightness for comparable illumination conditions by more than one order of magnitude.  

\section{Conclusion}
In this work, we have demonstrated the development and characterization of a laser-driven cold-field emission source tailored for ultrafast transmission electron microscopy. This pulsed photoemitter combines a small energy width of 360 meV with an exceptionally high peak brightness of over $6.7\times 10^{13}$ A/m$^2$sr, significantly advancing the capabilities of UTEM in probing ultrafast nanoscale dynamics. Finally, we note that in the current instrumental setting, few-Angstrom photoelectron spot sizes are only achievable for largely spatially coherent beams and are therefore requiring substantial filtering of the beam with high associated losses in electron currents. Settings with larger convergence angles would allow for similar spot sizes with less coherent beams at higher photocurrents assuming non-aberrated electron optics. Therefore, the implementation of laser-driven cold-field emitters in probe-corrected electron microscopes seems to be highly advantageous.

\section{Acknowledgement}
We acknowledge financial support by the Volkswagen Foundation as part of the Lichtenberg Professorship "Ultrafast nanoscale dynamics probed by time-resolved electron imaging" and funding by the German Science Foundation within the grant INST 184/211-1 FUGG. Further partial funding is provided by the Free State of Bavaria through the Lighthouse project "Free-electron states as ultrafast probes for qubit dynamics in solid-state platforms" within the Munich Quantum Valley initiative.

\newpage

\bibliography{bib}

\newpage

\setcounter{equation}{0}
\setcounter{figure}{0}
\setcounter{table}{0}
\setcounter{page}{1}
\makeatletter
\renewcommand{\theequation}{S\arabic{equation}}
\renewcommand{\thefigure}{S\arabic{figure}}

\noindent\textbf{\large{Supporting information for: \\ Laser-driven cold-field emission source for ultrafast transmission electron microscopy}}

\vspace{2cm}
\noindent \textbf{This supporting information includes:} \\ 
Additional information on the measurements of the photocurrent, longitudinal and transverse beam characterizations, technical information on the emitter vacuum and the electron emission model. 

\newpage

\subsubsection{Photocurrent measurements}
For characterizing the emitted photocurrent and its dependence on the incident wavelength and extraction field without the interference from apertures within the electron microscopy column, the laser-driven cold-field emitter gun was mounted on a separate ultrahigh-vacuum chamber equipped with a chevron microchannel plate assembly (MCP) with an attached phosphor screen. The electron gun is operated by a programmable high-voltage source identical to the source used for TEM operation. Due to X-ray safety reasons, the operation of the gun on the characterization chamber was limited to electron energies of 50 kV. The electric field environment around the tip and subsequent electrostatic focusing was similar to the case of 200-kV acceleration potential, since part of the acceleration section was electrically shorted in this mode.   
For high extraction fields at the emitter, the tunnel emission current is sufficiently larger so that it can be measured directly at the front plate of the MCP with a sensitive femtoampere meter (Keithley 6430 Sourcemeter with preamplifier). In this case a positive potential of 20 V is applied to the front plate in order to collect emitted secondary electrons. To quantitatively measure the total electron current transmitted through the CFEG optics in photoemission mode, the smaller achievable  current needs to be amplified by the MCP and a camera outside of the vacuum chamber records the image on the phosphor screen. To calibrate the camera, we conducted measurements at intermediate electron currents. At these levels, the camera and MCP were not saturated, but the current was still detectable by the femtoampere meter. This allowed us to fit the integrated image intensity recorded by the camera with the actual electron current measured by the femtoampere meter, ensuring an accurate calibration of the system. All measurements shown in Fig.~2 of the main manuscript, were recorded with a setting of the electrostatic focusing lens voltage within the emitter optimized for maximum transmission.

\subsubsection{Emitter Vacuum}
The vacuum within the acceleration section of the electron gun is measured with a hot cathode gauge, yielding a value of below $6 \cdot 10^{11}$ mbar. Close to the emitter tip additional non-evaporatable getter (NEG) pumps further decrease the local pressure $P$. For estimating the local pressure, we utilized an approach by Cho et al.$^1$ who have compared for a W(310) emitter the quality of the vacuum  with the lifetime of field emission, developing an empirical correlation $P \propto \frac{1}{\tau_{50}}$. Assuming a similar composition of the residual gas as in this work and considering a 50\%-damping constant $\tau_{50}$ of at least 900 minutes (cf. Fig. 1(g) of the main manuscript), we obtain a local (effective) pressure at the emitter of below $2 \cdot 10^{-11}$ mbar.

\subsubsection{Electron emission model}
The calculation of the field emission current is based on the work of Forbes$^2$, where the current density is approximated by the Fowler-Nordheim equation$^{3,4}$ $J_F=\frac{Aa}{\phi t^2}F^2 \mathrm{exp}[-vb\phi^{3/2}/F]$ with the emission area $A$ and the constants $a=e^3 / 8 \pi h$ and $b=\frac{4}{3} \sqrt{2m_e}/eh$. The factors $t$ and $v$ can be approximated as $v(y)\approx1-y^2+(1/3)y^2\mathrm{ln}y$ and $t(y)\approx 1+(1/9)[y^2-y^2\mathrm{ln}y]$, and depend on a single variable $y=\sqrt{e^3F/4\pi\epsilon_0}/\phi$. \\
The photoemission current is calculated based on Dowell and Schmerge$^5$ with $J_{\mathrm{photo}}=N_p \mathrm{QE}$ with the number of Photons $N_p$ and the quantum efficiency $QE=\frac{1-(\omega)}{2}\frac{(E_F+\hbar\omega)}{(2\hbar\omega)}[1-\sqrt{\frac{E_F+\phi}{E_F+\hbar\omega}}]^2$. Here $R(\omega)$ is the cathode optical reflectivity.

\subsubsection{Spectral and temporal electron puls characterization}
The characterization of the electron energy spectra was performed using a post column energy filter (CEOS CEFID) and a CMOS camera (TVIPS XF-416). The specified energy resolution of the CEFID is less than 100 meV (FWHM), well below the expected energy width of the photoelectron beam. At an energy dispersion of 16 eV/chip-width, the resulting energy resolution per pixel is 3.9 meV. In order to determine the electron pulse length, an electron-photon cross-correlation is performed. Here the electron pulse and a second light pulse arriving on the sample (800-nm wavelength, 169 fs pulse duration) are delayed in time using a motorized linear stage with a typical accuracy of 11 fs.

\subsubsection{Brightness determination}
The photoelectron caustics and brightness measurements were performed with the electron gun placed on the Regensburg UTEM. For imaging the electron focal spot size a nominal image magnification of either 1.2 or 2M is chosen. The image is recorded with a CMOS camera (TVIPS XF-416, 4096 x 4096 pixels) mounted in front of the energy filter. The projected pixel size was calibrated to be 0.118 $\mathrm{\mathring{A}}$/pixel (1.2M) and $0.056 \mathrm{\mathring{A}}$/pixel (2M), respectively. 
The angular distributions were measured with a direct electron detector (CheeTah T3, Amsterdam Scientific Instruments, 512 x 512 pixels, sensor size of 28 mm x 28 mm) mounted close to the projection lens. The detective quantum efficiency (DQE) at zero spatial frequency is close to unity for the CheeTah T3 at high electron energies$^6$ ensuring that the total electron current can be accurately determined by counting the detected electron event clusters.


\subsubsection{Brightness characterization for further settings of the condenser system}
\begin{table}[]
\resizebox{\textwidth}{!}{%
\begin{tabular}{@{}c|cccc|cccccc@{}}
\multicolumn{1}{l|}{} & \begin{tabular}[c]{@{}l@{}}Optical\\ Power\\ (mW)\end{tabular} & Spot & Angle & CLA & \begin{tabular}[c]{@{}l@{}}Width \\ (nm)\end{tabular} & \begin{tabular}[c]{@{}l@{}}$\alpha$ \\ (mrad)\end{tabular} & \begin{tabular}[c]{@{}l@{}}I\\ (fA)\end{tabular} & \begin{tabular}[c]{@{}l@{}}$\varepsilon$\\ (nm mrad)\end{tabular} & \textit{K} & $\begin{array}{c}
\text{Brightness} \\
(10^7 \mathrm{A/m}^2\mathrm{sr})\\
\left( \begin{array}{c}
     \text{Peak Brightness}  \\
     10^{13} (\mathrm{A/m}^2\mathrm{sr})
\end{array} \right)  \\
\end{array}$\\ [2ex]\midrule
\multirow{8}{*}{\begin{tabular}[c]{@{}c@{}}CW Photoemission\\ TEM Mode\end{tabular}} & 25 & 1 & 5 & 200 & 0.62 & 24.9 & 66.0 & 3.15 & 0.06 & 16.8    \\ [2ex]
 & 25 & 1 & 5 & 100  & 0.57 & 12.6 & 17.5 & 1.47 & 0.13 & 20.5    \\ [2ex]
 & 25 & 1 & 5 & 40 & 0.45 & 5.26 & 2.63 & 0.49 & 0.39 & 27.8   \\[2ex]
 & 25 & 1 & 5 & 10  & 1.08 & 2.01 & 0.14 & 0.45 & 0.43 & 1.77     \\[2ex]
 & 25 & 1 & 1 & 40 & 0.53 & 4.48 & 3.13 & 0.49 & 0.39 & 32.4    \\[2ex]
 & 25 & 1 & 3 & 40 & 0.50 & 4.99 & 3.30 & 0.52 & 0.38 & 31.5    \\[2ex]
 & 25 & 1 & 5 & 40& 0.45 & 5.40 & 3.05 & 0.49 & 0.39 & 31.5    \\[2ex]
 & 25 & 5 & 5 & 40 & 0.39 & 5.29 & 1.83 & 0.42 & 0.46 & 26.0    \\ [2ex]\midrule
\multirow{5}{*}{\begin{tabular}[c]{@{}c@{}}CW Photoemission\\ Probe Mode\end{tabular}}  & 25 & 1 & 2 & 40 & 0.68 & 3.75 & 2.73 & 0.52 & 0.37 & 25.46   \\ [2ex]
& 25 & 1 & 3 & 40 & 0.57 & 4.23 & 2.66 & 0.49 & 0.39 & 27.4    \\[2ex]
 & 25 & 1 & 5 & 40 & 0.31 & 8.09 & 2.76 & 0.51 & 0.38 & 27.3    \\[2ex]
 & 25 & 3 & 5 & 40 & 0.27 & 8.16 & 1.56 & 0.45 & 0.43 & 19.6   \\[2ex]
 & 25 & 5 & 5 & 40 & 0.21 & 8.22 & 0.63 & 0.36 & 0.54 & 12.5    \\[2ex]\midrule
\multirow{3}{*}{\begin{tabular}[c]{@{}c@{}}fs Photoemssion\\ TEM Mode\end{tabular}} & 1.04 & 1 & 1 & 40 & 3.20 & 4.64 & 1.30 & 3.07 & 0.07 & \begin{tabular}[c]{@{}l@{}}0.37\\ (5.16) \end{tabular} \\[2ex]
 & 0.3 & 1 & 1 & 40 & 1.52 & 4.65 & 0.39 & 1.46 & 0.14 & \begin{tabular}[c]{@{}l@{}}0.48\\ (6.69)\end{tabular} \\[2ex]
& 0.19 & 1 & 1 & 40 & 1.33 & 4.60 & 0.21 & 1.26 & 0.15 & \begin{tabular}[c]{@{}l@{}}0.35\\ (4.68)\end{tabular} \\ [2ex]\midrule
\multirow{3}{*}{\begin{tabular}[c]{@{}c@{}}fs Photoemission\\ Probe Mode\end{tabular}}  & 1.04 & 5 & 5 & 40 & 1.07 & 7.24 & 0.26 & 1.60 & 0.12 & \begin{tabular}[c]{@{}l@{}}0.26\\ (3.63) \end{tabular} \\[2ex]
 & 0.3 & 5 & 5 & 40 & 0.63 & 7.24 & 0.09 & 0.95 & 0.21 & \begin{tabular}[c]{@{}l@{}}0.25\\ (3.50)\end{tabular} \\[2ex]
 & 0.19 & 5 & 5 & 40 & 0.52 & 7.30 & 0.05 & 0.79 & 0.25 & \begin{tabular}[c]{@{}l@{}}0.20\\ (2.72)\end{tabular}
\end{tabular}
}
\caption{Transverse photoelectron beam properties for various condenser settings, operated in both CW photoemission and fs-pulsed photoemission mode. For a given set of beam parameters (Spot, Angle and CLA) the retrieved characteristics include the minimum focal spot size (FWHM), semi-convergence angle ($\alpha$), electron current ($I$), degree of coherence, and normalized brightness and normalized peak brightness. For CW excitation, the optical power was set to 25 mW, while for fs-pulsed operation, the optical power was chosen from 0.19 mW to 1.04 mW}
\end{table}
In addition to the discussion in the main part of the manuscript, Table 1 presents the experimental photoelectron beam properties for a broader range of condenser settings, showing the expected dependencies. For example, decreasing the CLA-aperture diameter from 200 to 40~$\mu$m in TEM/1/5 (for terminology see main part of the manuscript) reduces the angular width and beam current proportionally to the aperture diameter and area, respectively. Surprisingly, the resulting apparent beam brightness shows a maximum in the case of TEM/1/5/100 although in an ideal ray-optics imaging system, the brightness should not depend on aperture size. In our case, however, the beam brightness for the TEM/1/5/100 case is limited by spherical aberration, whereas for TEM/1/5/10 the beam approaches a fully coherent beam (relative coherence $K=0.43$) so that the diffraction-limit starts to govern the obtained focal spot size.\\
Increasing spot size settings, here for the measurement series Probe/1/5/40 to Probe/5/5/40, varies the spread of the electron beam across the condenser lens aperture. Thereby, the spatial coherence of the beam is increased at a fixed convergence angle. Remarkably, the degree of coherence $K=\xi_c/\sigma_x$, with the transverse coherence length $\xi_c=\hbar/m_{ec}\cdot \sigma_x/\varepsilon_{rms_x}$, reaches up to 54\%. Consequentially, the experimental electron focal spot size decreases from 3.05~\AA ~to 2.12~\AA. As observed above, the photoelectron current scales with the optical intensity on the emitter apex. For the low-current regime in which CW photoemission operates (estimated emission currents at the tip below 300 pA for 25-mW optical power and 3\% transmission), no Coulomb interactions between electrons in the CW photoelectron are expected. Correspondingly, the CW beam emittance is found to be independent of the optical intensity on the emitter and the beam brightness scales linearly with the optical intensity, as displayed in Fig. 4(f) of the main manuscript.

\newpage
\noindent \textbf{\Large{References}}
\begin{enumerate}
\item Cho, B.; Shigeru, K.; Oshima, C. W(310) cold-field emission characteristics reflecting the
vacuum states of an extreme high vacuum electron gun.\textit{ Review of Scientific Instruments}
\textbf{2013}, \textit{84}, 013305.
\item Forbes, R. G. Simple good approximations for the special elliptic functions in standard
Fowler-Nordheim tunneling theory for a Schottky-Nordheim barrier. \textit{Applied Physics
Letters} \textbf{2006},\textit{ 89}.
\item Fowler, Ralph, Howard; Nordheim, L. Electron emission in intense electric fields. \textit{Pro-
ceedings of the Royal Society of London. Series A, Containing Papers of a Mathematical
and Physical Character} \textbf{1928}, \textit{119}, 173–181.
\item Murphy, E. L.; Good, R. H. Thermionic Emission, Field Emission, and the Transition
Region. \textit{Physical Review} \textbf{1956}, \textit{102}, 1464–1473.
\item Dowell, D. H.; Schmerge, J. F. Quantum efficiency and thermal emittance of metal
photocathodes. \textit{Physical Review Special Topics - Accelerators and Beams} \textbf{2009}, \textit{12}.
\item van Schayck, J. P.; Zhang, Y.; Knoops, K.; Peters, P. J.; Ravelli, R. B. G. Integration of
an Event-driven Timepix3 Hybrid Pixel Detector into a Cryo-EM Workflow.\textit{ Microscopy
and Microanalysis} \textbf{2023}, \textit{29}, 352–363.
\end{enumerate}

\end{document}